\documentclass{article}
\usepackage{graphicx}

\newcommand{\E}[1]{\cdot 10^{#1}}
\newcommand{\grad}{^{\!\rm o \,}}

\def\beq{\begin{equation}}
\def\eeq{\end{equation}}
\def\beqn{\begin{eqnarray}}
\def\eeqn{\end{eqnarray}}
\def\nbeqn{\begin{eqnarray*}}
\def\neeqn{\end{eqnarray*}}

\def\bite{\begin{itemize}}
\def\eite{\end{itemize}}
\def\bdes{\begin{description}}
\def\edes{\end{description}}
\def\bfig{\begin{figure}}
\def\efig{\end{figure}}

\def\Re{\,{\mbox{Re}\,}}
\def\Im{\,{\mbox{Im}\,}}
\def\pr{\,\prime}

\def\ApO{\it Appl.~Opt.}
\def\ApP{\it Appl.~Phys.}
\def\PL{\it Phys.~Lett.}

\begin{document}
\begin{titlepage}
\vspace*{2.8cm}
\begin{center}
{\Large \bf{Alignment procedure for the VIRGO Interferometer:}} \\[10pt]
{\Large \bf{experimental results from the Frascati prototype.}} \\[40pt]
{\large {D.Babusci\footnote{{\it E-mail}: babusci@ibmllv.lnf.infn.it}, 
H.Fang, G.Giordano, G.Matone, L.Matone and V.Sannibale\footnote{Present 
address: {\it {L.A.P.P. Chemin de Bellevue BP 110,74941 Annecy-le-Vieux 
Cedex, France}}}}} \\[15pt]
{\it {INFN~-~Laboratori Nazionali di Frascati, P.O. Box 13, 
I-00044 Frascati, Italy}} \\[5pt]
\end{center}
\vspace*{2.0cm}
\begin{center}
{\large {\bf Abstract}} \\[25pt]
\end{center}
A small fixed-mirror Michelson interferometer has been built
in Frascati to experimentally study the alignment method that
has been suggested for VIRGO. The experimental results
fully confirm the adequacy of the method. The 
minimum angular misalignment that can be detected in the present 
set-up is 10 nrad/$\sqrt{\mbox{Hz}}$. \\[250pt]
\end{titlepage}

\section*{Introduction}

The search for Gravitational-Waves (GW) with interferometric 
antennas (VIRGO, LIGO, GEO) aims at the detection of a GW-signal 
by measuring the relative motion that a GW induces between two 
widely separated test masses \cite{giaz,abra,geoe}. In all these 
antennas, the test masses are the mirrors suspended at the ends of
two long arms, perpendicular to each other and forming a Michelson
interferometer, with Fabry-Perot cavities in the VIRGO and LIGO cases.
Inside the detector, IR-light (1.064 $\mu$m) is split into two beams 
which travel down the arms and reflect from the mirrors at the two ends. 
On their return, the beams interfere on the detection plane and from 
the illumination changes of the photodiode one can infere the relative 
changes that have occurred to the interferometer arm lengths. However, 
all the optical elements of the interferometer undergo position 
and angle fluctuations that are mostly due to thermal and seismic 
noise. Any angular tilt leads to a reduction in the sensitivity of the
interferometer and in particular causes a variation of the cavity 
length that simulates a GW-signal. 

The specifications required for the angular stability are determined by 
the sensitivity level one is aiming at for the detection of a GW-signal. 
These can turn out to be very severe and thus the sensitivities required 
in monitoring the error signals can become an issue. As a part of the 
research and development program in support of the VIRGO project, 
we built a small fixed-mirror interferometer with Fabry-Perot 
cavities in the two arms. The aim of the work was only to focus on the 
experimental verification of the theory developed to compute the alignment 
signals and on the method to extract from them the angular position of 
each individual mirror.

\vspace{1cm}
\renewcommand{\theequation}{\arabic{section}.\arabic{equation}}
\setcounter{equation}{0}
\section{The Fabry-Perot alignment methods}

Two are the basic alignment methods that we considered and both of 
them are based on the phase modulation of the incoming beam. 
The first was developed at Caltech by D.Z.Anderson et al.\cite{ande} 
and the second at Glasgow by H.Ward et al.\cite{ward}. Both of them 
were originally designed to align one single Fabry-Perot (FP) 
but no attempt has ever been made to extend this procedure to the 
case of a complete Michelson Interferometer (MI). 

For a brief description of the two methods, let us consider a pure 
TEM$_{00}$ laser beam entering a Fabry-Perot cavity with a 
tilting angle $\Delta\alpha$ with respect to the cavity axis. 
In the limit where this angle is much smaller than the far field 
divergence of the beam $\alpha_{0} = \lambda/\pi w_0$, the cavity 
sees the input beam as a linear superposition of the TEM$_{00}$ 
distribution and the first off-axis mode TEM$_{10}$. Similarly, 
if the TEM$_{00}$ beam enters the cavity with a lateral displacement 
$\Delta x$ much smaller than the beam waist $w_0$, the cavity sees again 
the incoming beam profile as a linear superposition of TEM$_{00}$ 
and TEM$_{10}$.

The difference between these two cases is that the rotations 
lead to a coupling into the TEM$_{10}$ mode as the translations do, but 
with a $90^{\circ}$ phase shift. This means that a misalignment causes 
a coupling into the lowest-order off axis mode with a phase that depends 
upon the type of misalignment. Therefore the transverse field 
distribution seen by the cavity as a consequence of small walks off 
of the two terminal mirrors, can be always approximated by a linear 
combination of these two cavity modes
\beq
E\, \simeq \,C_{0}\;(\;U_{0}\,+\,CU_{1}\;)\;,
\label{eq:1.1}
\eeq
where $E$ is the normalized input field, and $U_{0}$ and $U_{1}$ are 
the usual orthonormal Hermite-Gaussian functions of $x,y$ associated
with the TEM$_{00}$ and TEM$_{10}$ modes. $C_{0}$ and $C$ are the coupling 
coefficients and in particular:
\bite
\item for a pure translation $\Delta x$:$~~~~C=\Delta x/w_{0}$;
\item for a pure rotation $\Delta\alpha$:$~~~~~~~C = i\,
\Delta\alpha/\alpha_{0}$; 
\eite
$C_{0}$ is always real and, since it is always very close to unity,
it has been assumed equal to one.

Therefore the basic idea of the methods is to find a way to 
detect the amplitude and phase of this induced TEM$_{10}$ component. 

\vspace{0.3cm}
\subsection{The Anderson method}

With this method, one chooses to phase-modulate the input beam at 
the frequency separation between TEM$_{00}$ and TEM$_{10}$, which
in a plano-concave cavity (length $L$, mirror radius $R$) is given by:
\beq
\Delta\nu\,=\,\frac{c}{2 \pi L} \arccos{\sqrt{1\,-\,L/R}}
\label{eq:1.2}
\eeq

The field amplitude for a beam of phase modulated light at the 
optical frequency $\omega$, has the form
\vskip -1mm
\beq
E = E_{0} e^{i \omega t} \Big{\{} J_{0}(m) + 
\sum_{l=1}^{l=\infty} J_{l}(m) \big{[}\, e^{i l \Omega t} + 
(-)^{l} e^{-\,i l \Omega t}\, \big{]} \Big{\}}\;,
\label{eq:1.3}
\eeq
where $\Omega/2\pi = \Delta\nu$ is the modulation frequency, $E_0$ is a 
constant real vector, and $J_{l}(m)$ is the Bessel function of 
order $l$ and phase modulation index $m$. In this notation, the 
physical electric field is obtained by taking the real part of 
the complex quantities. If the value of the modulation index $m$ is 
sufficiently small, only the first three terms in the expansion 
(\ref{eq:1.3}) can be retained and the expression of the electric 
field (\ref{eq:1.1}) reduces to:
\beq
E \simeq E_{0}\,(U_{0} + C U_{1})\,e^{i \omega t}\,\Big{\{}
J_{0}(m) + 2 i J_{1}(m) \sin{\Omega t} \Big{\}}\;.
\label{eq:8}
\eeq
By retaining only the resonant terms, namely the fundamental 
mode at the carrier frequency and the TEM$_{10}$ mode at the 
sideband frequency ($\omega+\Omega$), the resulting transmitted 
intensity exhibits a spatially dependent component modulated at 
the beat frequency $\Omega$ as follows ($I_{0}=\mid E_{0} \mid^2$):
\beq
I\,=\,\mid {\cal T}^2 \mid \,I_{0}\,\Big{\{} J^{2}_{0}\,U^{2}_{0}\,+
\,\mid C \mid^{2}\,J^{2}_{1}\,U^{2}_{1}\,+\,2J_0 J_1 U_0 U_1\,
\Big{(} \Re [C] \cos{\Omega t}\,+\,\Im [C] \sin{\Omega t} \,
\big{)} \Big{\}}\;,
\label{eq:1.5}
\eeq
where ${\cal T}$ is the FP's complex transmittivity on resonance 
\cite{babu}.

Since the Hermite-Gaussian functions are mutually orthogonal when 
integrated all over the space, the detection of the entire transmitted 
beam by a single photodiode results in a DC-signal. The correspondent 
DC-photocurrent is obtained by integrating the first and leading term 
of eq.(\ref{eq:1.5}) yielding
\beq
I_{dc}=\eta\,\frac{e \lambda}{h c}\,\mid {\cal T} \mid^2\,
I_{0}\,J^{2}_{0}(m) \;,
\label{eq:10}
\eeq
where $\eta$ is the quantum efficiency. On the contrary, a 
separate detection of each half of the transmitted beam, followed by 
electronic subtraction of the two photocurrents, yields a current 
signal given by
\beq
I_{diff} = 2 \sqrt{\frac{2}{\pi}}\,I_{dc}\,\frac{J_{1}}{J_{0}}\,
\Big{\{}\,\Re[C] \cos{\Omega t}\,+\,\Im[C] \sin{\Omega t}\,\Big{\}}\;.
\label{eq:11}
\eeq
This equation shows that the intensity component that is in-phase 
with the modulation ($\cos{\Omega t}$) is proportional to the 
translational error and the in-quadrature component ($\sin{\Omega t}$) 
is proportional to the angular alignment error. Thus, by demodulating 
the signal, one can obtain simultaneously and independently both 
the misalignment errors. With the same argument, a quadrant photodiode 
detector permits simultaneous detection of couplings to both vertical 
and horizontal off-axis modes.

\vspace{0.3cm}
\subsection{The Ward method}\label{sec1.2}

This alternative technique was suggested by R.Drever at Caltech 
and experimentally demonstrated by H.Ward at the University of 
Glasgow \cite{ward}. It is basically an extension of the method used
for the longitudinal locking (Pound and Drever) \cite{drev} and 
it uses the light which is reflected from the FP. The beam is phase 
modulated but the modulation frequency $\Omega$ is not to be equal 
to the frequency difference between fundamental and first transverse 
mode as it is in the Anderson method. 

Let us suppose that a plano-concave cavity is both laterally and angularly 
misaligned with respect to the incoming beam direction of the 
usual quantities 
$a = \Delta x/w_0$ and $\alpha = \Delta \alpha/\alpha_0$ 
respectively, with $\Delta x$ being measured at the waist position.
In the cavity frame the incoming beam is described by
\beq
E^{\pr}_{in} = E_{0} e^{i \omega t} \Big[ U_{0} + 
(a + i \alpha) U_{1} \Big] \Big[ J_{0} + 
2 i J_{1} \sin{\Omega t} \Big] \;,
\label{eq:33}
\eeq
where U$_0$ at the carrier frequency is the only resonant term,
and all the other terms are assumed to be completely off-resonance.
With the convention that the reflection from a mirror introduces a 
90 $\grad$ phase shift, the phase of a beam reflected from a 
Fabry-Perot is $- 90 \grad$ ($+ 90 \grad$) depending on whether the
beam is on(off)-resonance. Assuming a totally reflective end mirror,
the expression for the reflected beam in the cavity frame is
\beq
E^{\pr}_{ref} = -\,i E_{0} e^{i \omega t} \Big{\{}
J_{0} \Big[ U_{0}- (a + i \alpha) U_{1} \Big] - 2 i J_{1} 
\Big[ U_{0} + (a + i \alpha) U_{1} \Big] \sin{\Omega t} \Big{\}}\;.
\eeq
which, transformed into the incoming beam frame, reads as follws
\beq
E_{ref} = -\,i E_{0} e^{i \omega t} \Big[
J_{0} (U_{0}-2a U_{1}) - 2 i J_{1} 
(U_{0} + 2 i \alpha U_{1}) \sin{\Omega t} \Big]\;.
\label{eq:1.9}
\eeq

The idea of the method is to let this beam evolve freely in space 
and to consider that an additional term intervenes in this process. 
This term is the phase difference between the real Hermite-Gaussian beam 
and an ideal plane wave, given by the Guoy phase
\beq
\phi_{n,m}(z) = (n+m+1)\,\phi(z)\;, \qquad
\tan{\phi(z)} = \frac{\lambda z}{\pi w_{0}^2}\;.
\eeq
The indexes $n,m$ refer to the $(n,m)$-th order mode and $z$ is the 
propagation coordinate whose origin, $z=0$, is at the location of 
the beam waist. This means that different modes evolve differently 
and the two components $U_0$ and $U_1$ acquire the phase difference 
$\phi(z)$. The eq. (\ref{eq:1.9}) becomes:
\beqn
E_{ref}(z) = -\,i E_{0} e^{i (\omega t + \phi)}\!\!\!\!&\Big{\{}
&\!\!\!\! \Big[ J_{0} (U_{0} - 2a U_{1} \cos{\phi})\,-\,
4 \alpha J_{1} \cos{\phi} \sin{\Omega t} \big{]} \nonumber \\
\!\!\!\!& & -\,2 i\, \Big[ a J_{0} U_{1} \sin{\phi} + 
J_{1} (U_{0} - 2 \alpha U_{1} \sin{\phi}) \sin{\Omega t} 
\Big]\,\Big{\}}\;.
\label{eq:36}
\eeqn

The intensity associated with this field depends upon the position 
where the detector is located. If the current difference between 
the two halves of a photodetector is taken at a given $z$-position, 
and the signal is demodulated at the frequency $\Omega$, the dominant 
component that will be detected is given by:
\beq
I_{diff}\:\propto \:-\:J_{0}J_{1}U_{0}U_{1} 
(a \sin{\phi} + \alpha \cos{\phi}) \sin{\Omega t}\;. 
\label{eq:37}
\eeq

Since the beam has a waist on M$_1$, $\sin{\phi}=0$ right behind 
M$_1$ and in this region the method is sensitive only to tilts. 
However, if the beam is let evolve, the angle $\phi$ goes to 
$\pi$/2 and so, at a very large distance from the waist, $I_{diff}$ 
becomes sensitive only to displacements. A more practical way to 
change $\phi$ is to employ an appropriate telescope, as discussed 
in detail in ref. \cite{ward}.

\vspace*{1cm}
\renewcommand{\theequation}{\arabic{section}.\arabic{equation}}
\setcounter{equation}{0}
\section{Beam evolution inside a complex interferometer}
\vspace*{0.2cm}

Let us consider a TEM$_{00}$ phase modulated beam $\Psi_{in}$
\beq
\Psi_{in} = U_0 \, e^{i \omega t}\,(J_{0} + 2 i J_{1} 
\sin{\Omega t})
\label{eq:psiin}
\eeq
impinging on a misaligned interferometer of a VIRGO-like structure 
(see fig.\ref{fig:itfform}) with the beam waist on M$_0$. In the 
following the "main frame" will always refer to this input beam.

\bfig[!ht]
\centering
\includegraphics[width=5in]{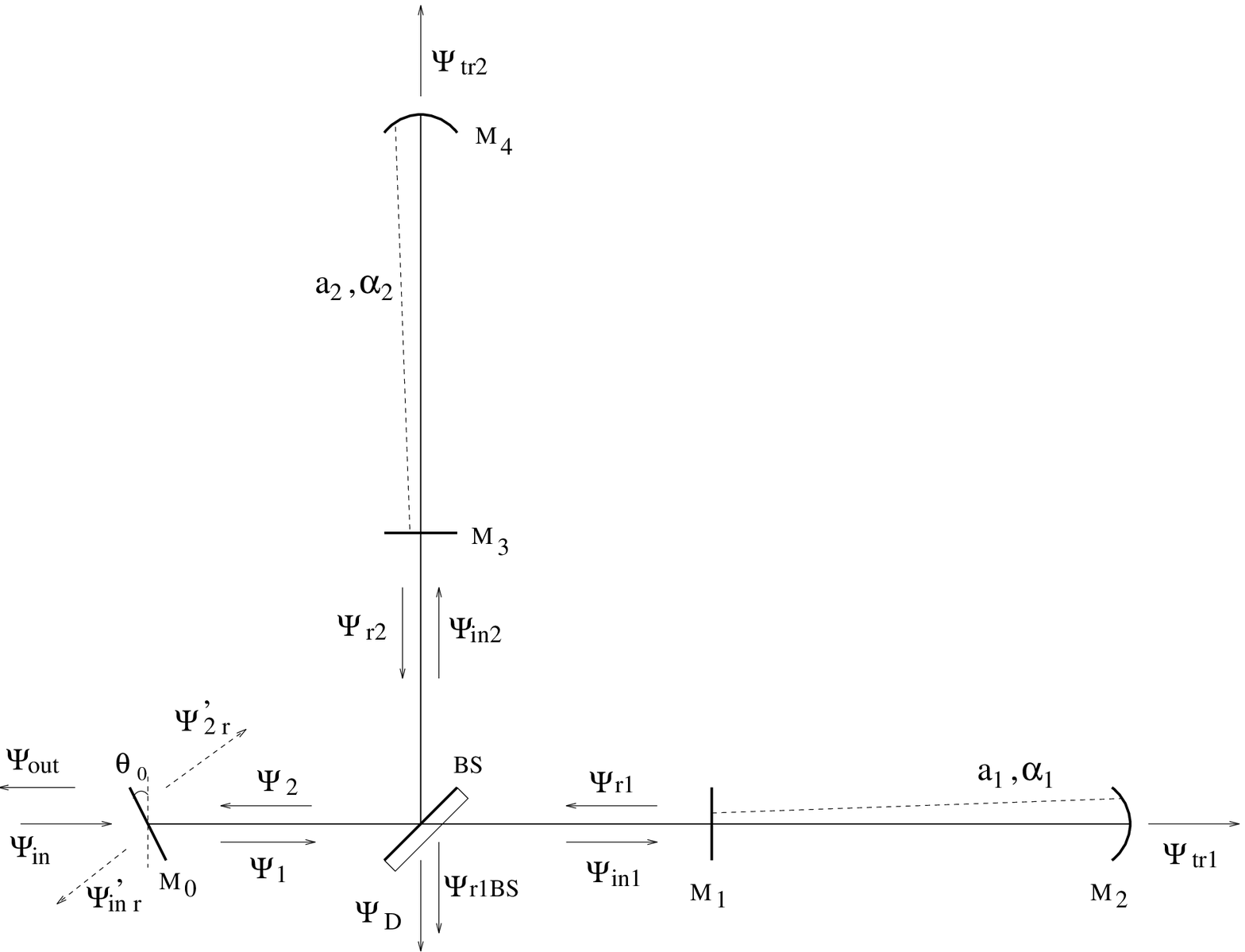}
\caption{Beams in a misaligned interferometer}
\label{fig:itfform}
\efig

In order to compute the expected alignment signals, we have 
developed a mathematical procedure that, under the following 
approximations and assumptions, leads to a fully analytical 
solution:
\bite
\item[1)] the beam mismatches, due to the difference
in length of the Michelson arms, and to the plane-plane configuration
of the recycling cavity (VIRGO case), have been neglected
\item[2)] all the misalignment angles are assumed to be much smaller than
the angular divergence of the beam
\item[3)] second or higher order terms in the misalignment angles have
been neglected in the expressions of the beam amplitudes
\item[4)] since, under the above approximations, no horizontal-vertical
coupling can be envisaged, the analysis has been limited to
one plane only
\item[5)] since in Virgo the angular noises that need to be corrected are
confined in the region below 1 Hz, the calculation has been performed in
a static condition, i.e. with all the optical elements being kept at their
misaligned positions.
\eite

Under the above mentioned conditions, all the beams running along or 
emerging from the interferometer, can be expressed as a linear combination 
of TEM$_{00}$ and TEM$_{10}$. Hence, each frequency component of the beam 
$\Psi_1$ (see fig. \ref{fig:itfform}) can be written in the following form
\beq
\Psi_1 = A\;e^{i\;\theta}\; \Big{[}U_0 + (a +
i\;\alpha)\;U_1\Big{]}
\label{eq:psi1}
\eeq
where $A$,$\theta$,$a$,$\alpha$ are the four quantities to be determined. 
Once this expression is given, all the other beams follow immediately, 
since the distances between the optical elements and the propagation
phases are all known. After the beam $\Psi_1$ has been transported to the 
two FPs, the reflected beams $\Psi_{r1}$ and $\Psi_{r2}$ are calculated 
according to a procedure similar to the one of sec.(1.2). These two 
reflected beams are then propagated back toward the beam splitter and 
their recombination $\Psi_{2}$ hits the recycling mirror M$_0$ which 
is supposed to be misaligned of the angle  $\theta_0$. The expression 
for the reflected beam in the main frame, takes the form
\beq
\Psi_{2r} = A^{*}\;e^{i\;\theta^{*}}\; \Big{[}U_0 + (a^{*} +
i\;\alpha^{*})\;U_1\Big{]}
\eeq
where the quantities $A^{*}, \theta^{*}, a^{*}, \alpha^{*}$ contain
$A, \theta, a, \alpha$, the propagation phases, and the angles of all
the mirrors. Finally, by imposing that the sum 
$\Psi_{2r} + t_0 \Psi_{in} = \Psi_{1}$, one obtains :
\beq
t_0 U_0 J_i \; + 
A^{*}\;e^{i\;\theta^{*}}\; \Big{[}U_0 + (a^{*} +
i\;\alpha^{*})\;U_1\Big{]}
 = A\;e^{i\;\theta}\; \Big{[}U_0 + (a + i\;\alpha)\;U_1\Big{]}
\label{eq:sum}
\eeq

By equalizing real and imaginary parts of U$_0$, U$_1$, eq.(\ref{eq:sum})
can be broken down into a system of four linear equations in the four
unknown $A, \theta, a, \alpha$. The solution of this system shows that 
$A$ and $\theta$ do not depend on the misalignment angles, while $a$ 
and $\alpha$ are given by:
\beq
a\,=\,\sum_{j = 0}^4\;b_j\,\theta_j \;, \qquad
\alpha\,=\,\sum_{j = 0}^4\;c_j\,\theta_j 
\eeq
where the $b_i, c_i$ coefficients depend on the geometrical and
optical characteristics of the interferometer (propagation phases,
mirrors transmittivities and reflectivities).

Once this procedure has been repeated for all the three frequency 
components, we have the full description of the beam inside the 
recycling cavity that can be propagated toward any position of 
interest, inside or outside the interferometer. Given a quadrant 
photodiode placed on an output beam, four signals can be detected: 
up-down and left-right differences demodulated in-phase and 
in-quadrature. In general, a signal $S^i$ from the $i$-th quadrant 
photodiode will be of the form:
\beq
S^i\,=\,\sum_{k = 0}^4\;C_k^i\,\theta_k
\label{eq:sig}
\eeq
This means that each signal will receive contributions from any 
misalignment and one has to find a way to extract the information 
relative to each of the five mirrors. The most common way of solving 
a problem of this kind is the $\chi^2$ procedure: having $n$ signals 
(with $n \geq 5$), the best estimate of the angles $\theta_j$ is 
found searching for the set of values leading to minimum $\chi^2$. 
In this case, given the simple structure of eq.(\ref{eq:sig}), this 
set of $\theta_j$-values can be obtained analytically, together with 
their uncertainties $\Delta \theta_j$, due to the presence of noise 
on the signals (shot-noise and electronic noise).

The reconstruction of the mirror angles can be optimized by choosing 
a sufficient number of quadrant photodiodes and by studying the 
dependence of their signal on the Guoy phase.

\vspace{1cm}
\renewcommand{\theequation}{\arabic{section}.\arabic{equation}}
\setcounter{equation}{0}
\section{The experimental apparatus}

The Frascati prototype is a small-scale, fixed-mirror Michelson 
interferometer in air, sufficient for us to concentrate on optical 
alignment and stability studies. The mechanical and optical apparatus 
lies on top of a TMC-optical table (1.5 m $\times$ 2.0 m), 
which is isolated from seismic ground motion with an air 
over-pressure spring system that couples it to the ground floor. 
A sound absorbing enclosure protects the whole assembly against the 
ambient acoustical noise. The interferometer is constructed with 
commercial mirror mounts and the space between the cavity mirrors is 
protected by plexiglass tubes to minimize the fluctuations in the 
refraction index induced by air circulation.

\vspace{0.3cm}
\subsection{The interferometer layout}

The layout is shown in fig. \ref{fig.1}. Light from a frequency 
stabilized Newport NL-1 He-Ne laser (0.5 mW at 632.8 nm) is phase 
modulated and focused with two positive lenses (L$_0$, L$_1$) 
into a beam waist of 5.9 $\cdot$ 10$^{-2}$ cm at the position of the 
recycling mirror (M$_0$), 1.94 m away from the laser output. 

\bfig[!ht]
\centering
\includegraphics[width=5in]{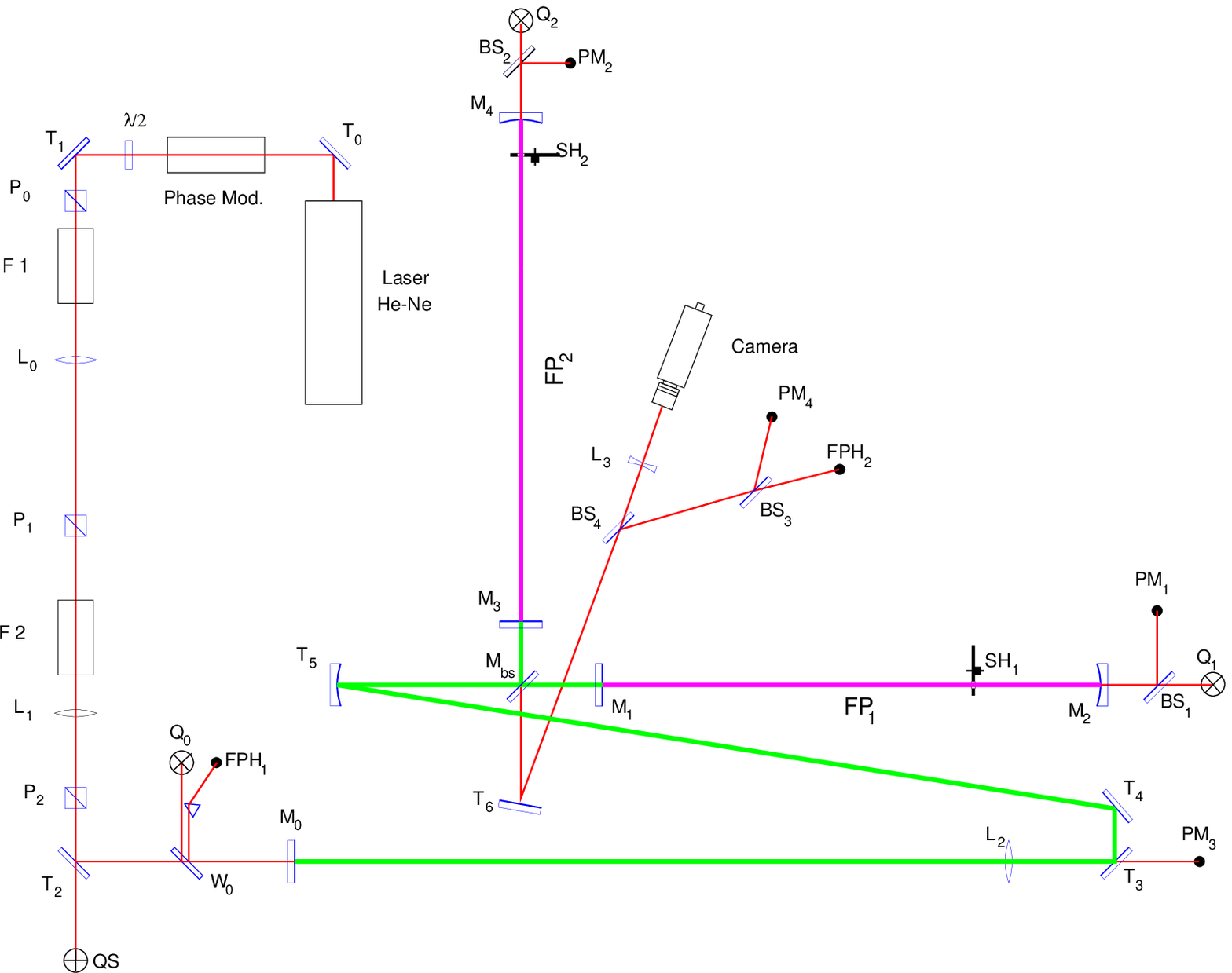}
\caption{Complete lay-out of the whole interferometer}
\label{fig.1}
\efig

The laser is isolated from spurious light feed-backs with a system of 
two Faraday rotators and two polarizing beam splitters. If only one 
of these two systems is used, the back reflection reduces to a 
fraction of 2.0 $\cdot$ 10$^{-5}$ of the incident power but becomes 
undetectable with a cascade of two. A third polarizing beam splitter 
(P$_2$) is added before the entrance of the interferometer to ensure 
linear polarization in the horizontal plane for both the beams 
directed to and reflected from the interferometer. A quadrant photodiode 
(QS) is placed on the beam transmitted from the mirror T$_2$: low 
frequency movements of the beam directed to the interferometer, due 
mainly to angular jitters of the laser output, are so detected and 
corrected, acting in a feedback loop on two piezos mounted on the 
mirror T$_0$.
 
To find room on the optical table, the recycling cavity is folded 
over a zig-zag sequence of two flat mirrors (T$_3$, T$_4$), 
a positive lens (L$_2$) and a curved mirror (T$_5$) with a total 
optical path of 2.68 m. The beam waist at the position of the two 
flat mirrors (M$_1$, M$_3$) is 7.2 $\cdot$ 10$^{-2}$ cm and is mode 
matched to the geometry of the two following FP-cavities. 
The FP mirrors (M$_{1,2}$ and M$_{3,4}$) and the recycling mirror 
(M$_0$) are mounted on Burleigh PZ-91 (piezo-mounting) to allow 
for fine longitudinal (2 nm/V) and angular adjustments (80 nrad/V 
horizontal, 92 nrad/V vertical). The mirror characteristics have been 
measured and the results are summarized in table I.

\begin{center}
Table I
\end{center}
\begin{center}
\begin{tabular}{||c|c|c||}
\hline
 &  &  \\
 Mirror  &  Transmittivity $t^2$ ($\%$)  & Curvature Radius (m)   \\
 &  &  \\
\hline
\hline
  M$_0$  &  40.8 $\pm$ 0.3  &  $\infty$  \\
\hline
  M$_1$,M$_3$  &   10.1 $\pm$ 0.2  & $\infty$ \\
\hline
  M$_2$,M$_4$  &  0.2 $\pm$ 0.01 &  10  \\
\hline
  BS     &  50.3 $\pm$ 0.7  & $\infty$ \\
\hline
  T$_3$  &  (6.0 $\pm$ 1.0) $\E{-2}$  &  $\infty$  \\
\hline
\end{tabular}
\end{center}
\vspace*{0.3cm}

The piezo mounts have a frequency response that relates the applied 
voltage to the motion effectively executed by the mirror. This response 
is described by a transfer function which, in general, depends upon the 
exciting frequency and can only be determined with an experimental 
measurement. In the low frequency interval ($\le$ 10 Hz), where most 
of the mechanical noise is concentrated, the transfer function is 
definitely well-behaved. However, structures are visible in the region 
between 200 Hz and 500 Hz and are associated with the mechanical structure 
of the mirror holder. They have been efficiently damped by adding an extra 
load to the mount.

\vspace*{0.3cm}
\subsection{Fabry-Perot's cavities}
 
Both the Fabry-Perot cavities are 70.6 cm long and the separation 
between the longitudinal modes is $c/2L\,=\,$212.3 MHz. From 
eq.(\ref{eq:1.2}), the transverse-longitudinal separation is 
18.2 MHz. The line-shape profile has been measured with a photodiode 
located behind the terminal mirrors and reproduces the expected 
expression in Ref. \cite{babu}. The measurement of the width $\Gamma$ 
of the resonance yields the finesse
\beq
{\cal F}\,=\,\frac{c}{2 L \Gamma}\,=\,\pi\,\frac{\sqrt{r_1 r_2}}
{1 - r_1 r_2}\,=\,57.5\,\pm\,2.0
\eeq
and allows for a direct measurement of the product 
$r_1 r_2\,=\,0.947\,\pm\,0.002$, in good accordance with the 
reflectivity values reported in table I. The value of the power 
enhancement (the power stored into the cavity per unit incident 
power), the reflectivities and the transmittivities of the FPs have 
been measured at resonance and the results are shown in table II
together with the same values expected on the base of the mirror 
transmittivities quoted in table I. The departure from one of the 
sum $T^2 + R^2$ measures the total amount of the cavity losses. 
They turn out to be 7.5 $\%$ and 9.4 $\%$ for FP$_1$ and FP$_2$ 
and imply coating losses of 0.11 $\%$ and 0.15 $\%$, both well 
within specs for standard commercial mirrors.

\begin{center}
Table II
\end{center}
\vspace{0.1cm}
\begin{center}
\begin{tabular}{||c|c|c|c||}
\hline
 &  &  &   \\
 &  Reflectivity $R^2$ ($\%$)  &  Transmittivity $T^2$ ($\%$)  
& Power Enh. \\
 &  &  &    \\
\hline
\hline
  FP$_1$  &  85.8 $\pm$ 3.0  &  6.7 $\pm$ 0.2  &  32.8 $\pm$ 1.0 \\
\hline
  FP$_2$  &  84.3 $\pm$ 3.0  &  6.3 $\pm$ 0.2  &  31.1 $\pm$ 1.0 \\
\hline
  expected  &  92.6  &  7.4  &  36.2  \\
\hline
\end{tabular}
\end{center}
\vspace{0.1cm}

Both FP cavities have been longitudinally locked with a "dithering" 
technique. The end mirrors of FP1 and FP2 are set into longitudinal
oscillation with an amplitude of $\sim 1.3 \cdot 10^{-5} \cdot \lambda$
at a frequency of 7.0 kHz and 4.3 kHz respectively. By means of two 
analog dividers, the detectors PM$_{1,2}$ have been normalized to 
PM$_3$ which is proportional to the recycling cavity inner power. 
The resulting ratios are then sent to two lock-in amplifiers, with
reference 7.0 kHz and 4.3 kHz, whose outputs constitute the error
signals of the feedback loops. These signals are then integrated, 
amplified and, finally, applied to the piezos of the two FP's terminal 
mirrors M$_2$ and M$_4$. In the ``feed-back on'' condition, the power 
fluctuactions in the two cavities never exceed a small percentage of 
the stored power. 

\vspace{0.3cm}
\subsection{The recycling cavity}

The recycling cavity comprises the sequence of elements that run from 
M$_0$ to M$_{1,3}$. It is arranged in a plano-plano geometry where 
the two flat terminal mirrors are connected by the two focusing 
elements L$_2$ and T$_5$. The distances from the beam splitter to 
the initial FP mirrors have been made slightly different 
(M$_{BS}$ - M$_1$ = 11.9 cm, M$_{BS}$ - M$_3$ = 9.4 cm), in analogy
to what they will be in VIRGO, to increase the signal on the dark
fringe detector \cite{flam}. The total cavity loss $\alpha_{rc}$, 
inclusive of the lens L$_2$ and the three mirrors T$_3$, T$_4$, T$_5$, 
has been obtained by a direct measurement of the power lost in one 
trip from L$_2$ to T$_5$. Unfortunately measurements of this kind are 
affected by large errors because they always require to take 
differences between almost equal numbers. In our case we have obtained 
$\alpha_{rc} = ( 1.92\,\pm\,1.0\;) \%$.

The recycling cavity can be looked at as a FP cavity where the 
initial mirror is M$_0$ and the terminal mirror is the ensemble of 
the beam splitter and the two FPs. This ensemble can be considered 
as an equivalent mirror with an effective reflectivity defined as:
\beq
r_{eq}^2\,=\,\frac{1}{2}\,(R^2_{FP_{1}}\,+\,R^2_{FP_{2}})\,(1\,-\,
\alpha_{rc})^2\,(1\,-\,\alpha_{fr})\,=\,( 81.5 \pm 2.6 )\,\%
\eeq
where $\alpha_{fr}$ accounts for the power that is inevitably lost 
in the incomplete light extinction on the fringe detector. This 
power loss has been evaluated to be of the order of 1 $\%$. The 
enhancement factor for this cavity is expected to be
\beq
F_{pe}\,=\,\frac{\mbox{stored power}}{\mbox{incident power}}\,=\,
\frac{T_0^2}{(1 - r_{eq} r_0)^2}\,=\,4.4 \pm 0.4
\label{eq:Fpe}
\eeq
and it has been obtained by measuring the power transmitted from 
T$_3$ (or equivalently by FP$_{1,2}$) with and without the insertion 
of M$_0$. The experimental result has been determined to be 
4.8 $\pm$ 0.3 in good agreement with eq.(\ref{eq:Fpe}). 

Two fast photodiodes were used in two feedback loops to keep the 
recycling cavity on resonance (FPH$_1$) and to maintain the dark 
fringe condition (FPH$_2$). The demodulated signal from FPH$_1$ 
was used to control the M$_0$ position, while the feedback from 
FPH$_2$ acted on both the FP$_2$ mirrors, thus constraining
the dark fringe without perturbing the status of FP$_2$.

\vspace{1cm}
\renewcommand{\theequation}{\arabic{section}.\arabic{equation}}
\setcounter{equation}{0}
\section{Experimental Results}
\vspace*{0.2cm}

According to the analytical simulation described in sec. 2 (for more 
details see ref. \cite{luca}) small mirror misalignments do not prevent 
the field amplitude from reaching an equilibrium condition that retains 
the information on each individual mirror position. This means that the 
demodulated signals from a quadrant photodiode placed on any of the beams 
leaving the interferometer are sensitive to the misalignment status of all 
mirrors, either in their in-phase or in-quadrature components, or in both. 
Furthermore, since each of these components can be analytically expressed 
in terms of the individual mirror misalignments, with a sufficient number
of measurements the alignment status of the whole interferometer can be 
fully determined. To better achieve this, the modulation frequency of the 
laser beam has to coincide with the TEM$_{00}$ $-$ TEM$_{10}$ frequency 
separation of the FPs (18.2 MHz in our case), since in this way the beams 
transmitted from the FPs have the maximum sensitivity to misalignments.  

In the experimental test we conducted, we set into oscillation M$_0$, 
M$_1$, M$_2$, M$_3$, M$_4$ one at a time, at the fixed frequency of 
3 Hz and with an amplitude of approximately 200 nrad. The signals 
seen by Q$_0$, Q$_1$ and Q$_2$ were each time demodulated at 
0 $\grad$ and 90 $\grad$ in order to construct the two quantities 
\beq
A (\mbox{M}_i)\,=\,\sqrt{I^2 (\mbox{M}_i) + Q^2 (\mbox{M}_i)}\;\qquad 
\varphi (\mbox{M}_i)\,=\,\arctan{\frac{I (\mbox{M}_i)}{Q (\mbox{M}_i)}}\;.
\label{eq:AFi}
\eeq
where $I(\mbox{M}_i)$ and  $Q(\mbox{M}_i)$ are the in-phase and 
in-quadrature signals generated by the $i$-th mirror.

Since the angle/voltage calibrations are the same for all the piezos in 
the system, it is quite appropriate to present the results normalized to 
one of them. Tables III,IV,V present the comparison between the experimental 
and expected values of these ratios for $Q_0$, $Q_1$ and $Q_2$, respectively.

\vspace*{2mm}
\begin{table}[!ht]
\begin{center}
Table III : Beam reflected from M$_0$ for a fixed value of 
$\Phi_G/\pi$: comparison between the experimental and theoretical 
values for the quantity $R_{ij} = A(\mbox{M}_i)/A(\mbox{M}_j)$ and 
$\Delta \varphi_{ij} = |\,\varphi(\mbox{M}_i) - \varphi(\mbox{M}_j)\,|$.
\end{center}
\begin{center}
\begin{tabular}{||c||c|c||c|c||c|c||}
\hline 
\hline
 & & & & & & \\
 $\Phi_G/\pi$ = 0.1063  & $R_{12}$  &  $\Delta \varphi_{12}$ (deg) &  
 $R_{20}$  &  $\Delta \varphi_{20}$ (deg) &  $R_{30}$  &  
 $\Delta \varphi_{30}$ (deg) \\ 
 & & & & & & \\
\hline 
\hline
 Expt.  &  1.45 $\pm$ 0.08  &  174.3 $\pm$ 3.0  &  
 1.33 $\pm$ 0.06  &  1.2 $\pm$ 1.6  & 
 1.33 $\pm$ 0.07  &  179.3 $\pm$ 3.1 \\
 & & & & & & \\
 Theor.  &  1.22  &  179.0  &  1.26  &  1.1  & 
 1.11  &  179.1 \\
\hline 
\hline
\end{tabular}
\end{center}
\end{table}

\begin{table}[!ht]
\begin{center}
Table IV : Beam transmitted from FP$_1$.
\end{center}
\begin{center}
\begin{tabular}{||c||c|c||c|c||c|c||}
\hline 
\hline
 & & & & & & \\
 $\Phi_G/\pi$ = 0.0415  &  $R_{02}$  &  $\Delta \varphi_{02}$ (deg) &  
 $R_{12}$  &  $\Delta \varphi_{12}$ (deg)  &  $R_{42}$  &  
 $\Delta \varphi_{42}$ (deg) \\ 
 & & & & & & \\
\hline 
\hline
 Expt.  &  0.40 $\pm$ 0.02  &  45.1 $\pm$ 3.0  &
 0.83 $\pm$ 0.03  &  173.8 $\pm$ 2.2  & 
 0.8 $\pm$ 0.1  &  25.2 $\pm$ 6.5 \\
 & & & & & & \\
 Theor.  &  0.395  &  49.6  &  0.835  &  174.7  & 
 0.7  &  23.0 \\
\hline 
\hline
\end{tabular}
\end{center}
\end{table}

\begin{table}[!hb]
\begin{center}
Table V : Beam transmitted from FP$_2$ for two values of $\Phi_G/\pi$
\end{center}
\begin{center}
\begin{tabular}{||c||c|c||c|c||}
\hline
\hline 
 & \multicolumn{2}{c||}{$\Phi_{G}/\pi$ = 0.0965} & 
   \multicolumn{2}{|c||}{$\Phi_{G}/\pi$ = 0.4144} \\
\cline{2-5}
 & & & & \\
 & Expt. & Theor. & Expt. & Theor. \\
\hline 
\hline
 & & & & \\
 $R_{02}$   & 0.52 $\pm$ 0.02 & 0.57 & 0.56 $\pm$ 0.01 & 0.55 \\
\hline
 & & & & \\
$\Delta \varphi_{02}$ (deg) & 10.1 $\pm$ 2.0 & 26.7 & 
15.5 $\pm$ 1.4 & 24.2 \\
\hline 
\hline
 & & & & \\
$R_{12}$ & 0.77 $\pm$ 0.02 & 0.91 & 0.9 $\pm$ 0.02 & 0.895 \\
\hline
 & & & & \\
$\Delta \varphi_{12}$ (deg) & 
183.8 $\pm$ 1.4 & 179.2 & 176.7 $\pm$ 1.0 & 179.2  \\
\hline 
\hline
 & & & & \\
$R_{32} $ & 1.28 $\pm$ 0.03 & 1.2 & 0.94 $\pm$ 0.02 & 1.17 \\
\hline
 & & & & \\
$\Delta \varphi_{32}$ (deg) & 146.6 $\pm$ 1.27 & 152.4 & 
137.6 $\pm$ 1.0  & 149.2  \\
\hline 
\hline
 & & & & \\
$R_{42}$ & 1.2 $\pm$ 0.03 & 1.43 & 1.18 $\pm$ 0.01 & 1.39 \\
\hline
 & & & & \\
$\Delta \varphi_{42}$ (deg) & 31.9 $\pm$ 1.1 & 22.4 & 
35.3 $\pm$ 0.6 & 24.6 \\
\hline 
\hline
\end{tabular} 
\end{center}
\end{table}

As shown in the tables, we have selected different values for the 
Guoy phases, either by choosing the appropriate position of the 
quadrant photodiode, or by inserting a lens between it and the output 
mirror, at whose position we define the zero of the Guoy phase. The 
quoted errors are only computed from the measured noise fluctuations. 
In all the cases the agreement appears extremely satisfactory.

The effect of the evolution of the Guoy phase has been directly 
tested on the beam reflected from M$_0$. As indicated at the end 
of section \ref{sec1.2}, we have kept Q$_0$ fixed at about 60 cm 
from M$_0$, and placed a lens with $f = +$ 20 cm at different 
distances from Q$_0$, thus obtaining values for the total Guoy 
phase in the range (0.11 $\div$ 0.94) $\pi$ rad. The behaviours of both 
$A(\mbox{M}_2)/A(\mbox{M}_0)$ and $\varphi(\mbox{M}_2) - 
\varphi(\mbox{M}_0)$, expressed in terms of the calculated Guoy 
phase $\Phi_G$, are shown in fig.\ref{fig.3}. 
\bfig[!ht]
\centering
\includegraphics[width=4.5in]{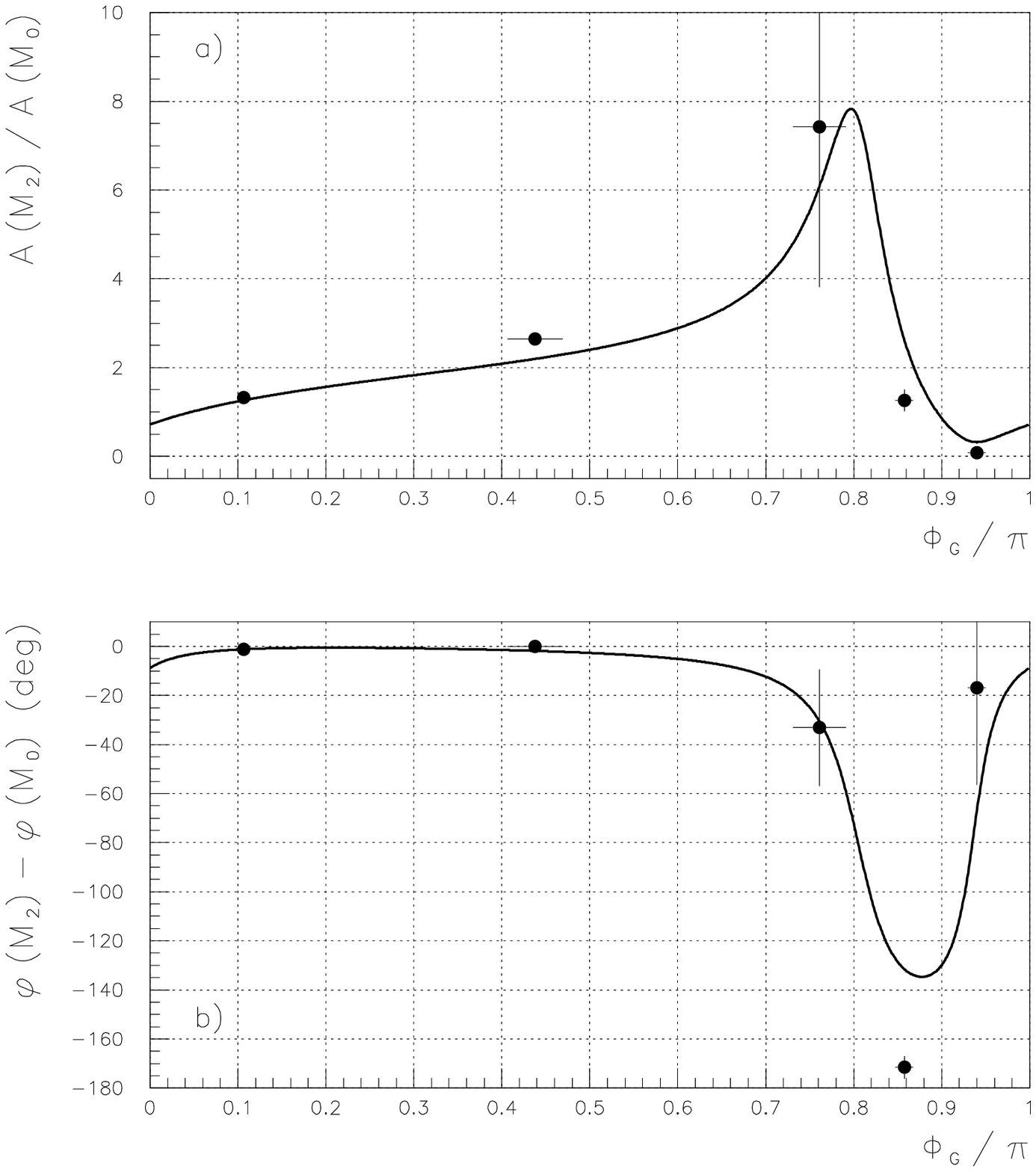}
\caption{Comparison between theory and measurements for the ratio 
$A (\mbox{M}_2)/A (\mbox{M}_0)$ (a) and the phase difference 
$\varphi (\mbox{M}_2) - \varphi (\mbox{M}_0)$ (b) expressed as 
a function of the Guoy phase $\Phi_G$.}
\label{fig.3}
\efig

The expected curve for $A(\mbox{M}_2)/A(\mbox{M}_0)$ peaks quite 
sharply at $\Phi_G/\pi\,=$ 0.8 and the experimental data follow 
this shape with a reasonable agreement. The smallness of the 
$A (\mbox{M}_0)$ value at the peak position makes the error bar 
on the ratio unusually large. The same observation holds for 
$\varphi(\mbox{M}_2) - \varphi(\mbox{M}_0)$ as well. By looking 
at the two figures, one could argue that there is a possible small 
systematic shift, of approximately 9 degrees, between theory and 
experiment. While there are a lot of possible explanations for that, 
one has to keep in mind that the theoretical curves have been 
computed using the measured values for distances, transmittivities etc.,
and thus they can be affected by systematic uncertainties as well.

The validity of this theoretical approach can be further appreciated 
in the following way. Starting from a coarse pre-alignment condition, 
one can read the residual error signals given by the three photodiodes 
and elaborate the corrections to be applied to each individual mirror 
for the fine alignment of the system. The sequence of the power levels 
that the dark fringe detector reads, when the corrections are applied 
one after the other, is presented in fig.(\ref{fig.4}). With this
correction procedure the light extinction on the dark fringe improves 
of more than a factor 3, reaching a value of 0.4 $\%$ of the total 
power stored into the cavity.
\bfig[!ht]
\centering
\includegraphics[width=4in]{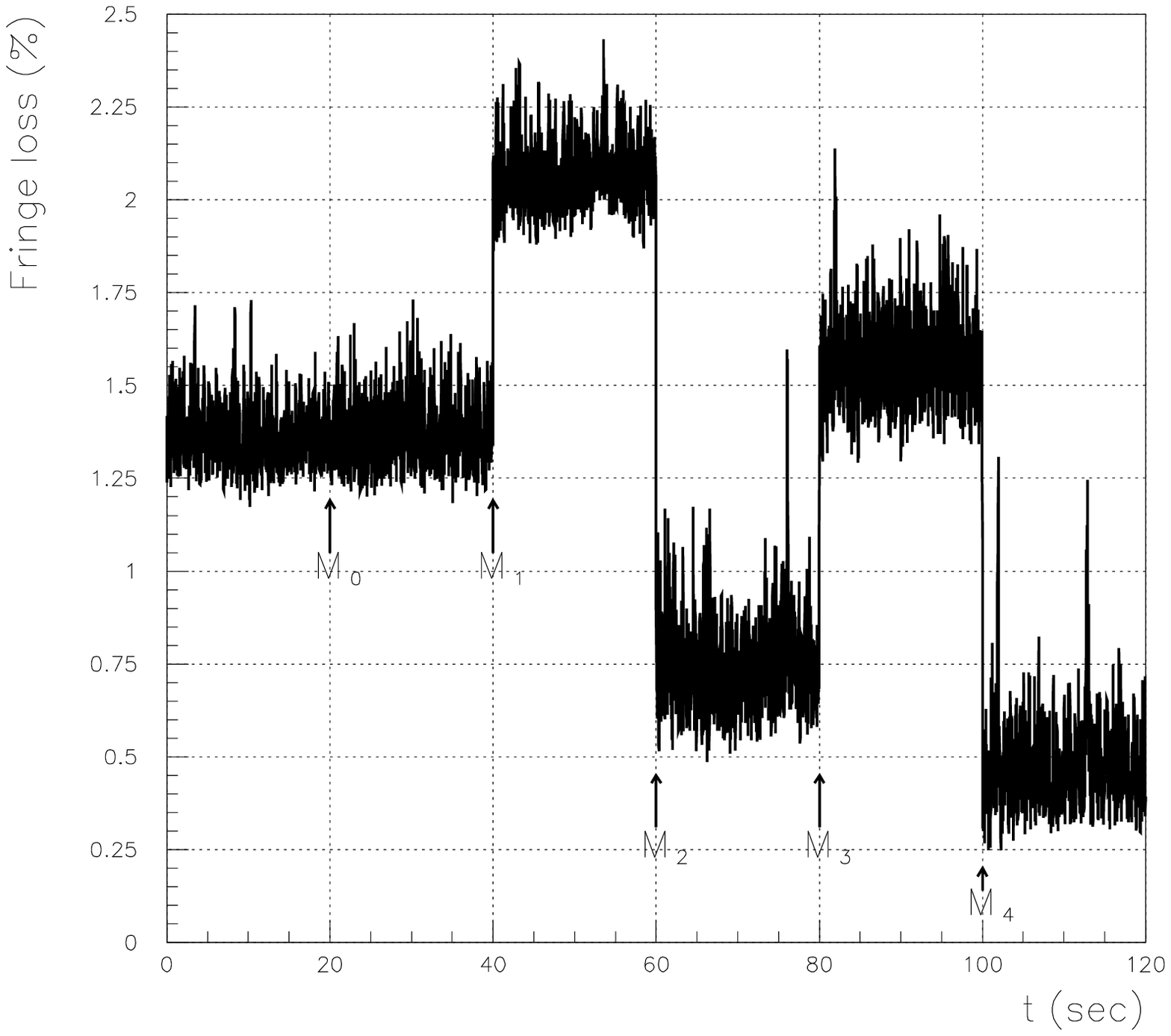}
\caption{Fringe loss (the ratio between the power on the fringe 
detector and the power in the recycling cavity) when the calculated 
corrections are applied in sequence to $\mbox{M}_{0,1,2,3,4}$.}
\label{fig.4}
\efig

In conclusion, we have demonstrated that even with a very low 
average power in the photodiodes, (1 $\div$ 30) $\mu$W, a small phase
modulation index (0.25 $\%$ of the total power in each sideband)
and a longitudinal locking scheme that could be largely improved, 
it is possible to accurately align a complex interferometer. Indeed, 
under these circumstances, the average noise level on the demodulated 
signals from the quadrant photodiodes is equivalent to about 10 
nrad/$\sqrt{\mbox{Hz}}$ or 4.0 $\cdot 10^{-5}\,\alpha_0/\sqrt{\mbox{Hz}}$.
Therefore, since this angle is the minimum angle that can be detected in 
our set-up, this quoted value represents the lowest limit to our 
present alignment capability. Finally, since the VIRGO sensitivity to 
GW-detection requires the angular noise to be reduced below 
1 $\mu$rad/$\sqrt{\mbox{Hz}}$ \cite{virg}, we claim that our 
present results already satisfy this requirement.

\vspace*{1.0cm}
\section*{Acknowledgments}
We want to express our appreciation to the VIRGO-Frascati technical
team, E. Cima, M. Iannarelli and E. Turri, for the great contribution
of work and assistance they gave us during the construction and
operation of the interferometer.

\end{document}